\documentclass[aps,pre,twocolumn]{revtex4}
\usepackage{graphicx}
\usepackage[dvips]{color}

\begin{document}

\title{Equilibrium free energies from fast-switching trajectories with
large time steps}

\author{Wolfgang Lechner, Harald Oberhofer, and Christoph Dellago}
\affiliation{Faculty of Physics, University of
Vienna, Boltzmanngasse 5, 1090 Vienna, Austria}

\author{Phillip L. Geissler} 
\affiliation{Department of Chemistry,
University of California at Berkeley, 94720 Berkeley, CA, USA}

\date{\today}

\begin{abstract}
Jarzynski's identity for the free energy difference between two
equilibrium states can be viewed as a special case of a more general
procedure based on phase space mappings.  Solving a system's equation
of motion by approximate means generates a mapping that is perfectly
valid for this purpose, regardless of how closely the solution mimics
true time evolution.  We exploit this fact, using crudely dynamical
trajectories to compute free energy differences that are in principle
exact.  Numerical simulations show that Newton's equation can be
discretized to low order over very large time steps (limited only by
the computer's ability to represent resulting values of dynamical
variables) without sacrificing thermodynamic accuracy.  For computing
the reversible work required to move a particle through a dense
liquid, these calculations are more efficient than conventional fast
switching simulations by more than an order of magnitude.  We also
explore consequences of the phase space mapping perspective for
systems at equilibrium, deriving an exact expression for the
statistics of energy fluctuations in simulated conservative systems.
\end{abstract}

\maketitle

\section{Introduction}

The maximum work theorem, a consequence of the second law of
thermodynamics, states that the amount of work performed {\em by} a
thermodynamic system during a transformation from a specific initial
state $A$ to a specific final state $B$ is 
less than energy difference between the two states \cite{CALLEN}. The
work $W$ is maximum and equal to the free energy difference, or {\em
reversible work}, if the transformation is carried out
reversibly. Equivalently, the average work performed {\em on} a system
during such a transformation is bounded from below by the free energy
difference $\Delta F$,
\begin{equation}
\langle W \rangle \geq \Delta F.
\label{equ:Clausius}
\end{equation}
The notation $\langle \cdots \rangle$ implies an average over many in
general irreversible transformations initiated 
in an equilibrium state.  (For macroscopic systems every individual
transformation will require the same amount of work but for small
systems work fluctuations occur.)  Remarkably, the inequality
(\ref{equ:Clausius}) can be turned into an equality by considering
exponential averages \cite{jarz},
\begin{equation}
\label{equ:jarzynski}
\exp(-\beta \Delta F)=\langle \exp (-\beta W)\rangle.
\end{equation}
where $\beta=1/k_{\rm B}T$ is the inverse temperature and $k_{\rm B}$
is Boltzmann's constant. This identity, proven by Jarzynski
\cite{jarz} and later by Crooks \cite{gavin_jstatphys} under very
general conditions, relates the statistics of irreversible work to
equilibrium free energy differences.

The Jarzynski identity can be used to calculate free energy
differences in computer simulations of molecular systems
\cite{jarz_efficiency,HUMMER,SUN,YTREBERG,VOTH}. In statistical
mechanical terms the free energy difference $\Delta F=F_B-F_A$ between a
system at temperature $T$ with Hamiltonian ${\cal H}_B(x)$ and another
one at the same temperature with Hamiltonian ${\cal H}_A(x)$ is given
by
\begin{equation}
\Delta F = -k_{\rm B}T \ln \frac{\int dx \exp \{-\beta {\cal
H}_B(x)\}}{\int dx \exp \{-\beta {\cal H}_A(x)\}}=-k_{\rm B}T \ln
\frac{Q_B}{Q_A},
\end{equation}
where the integration extends over the entire phase space and $x=\{q,
p\}$ includes the positions $q$ and momenta $p$ of all particles. In
the above equation, $Q_A$ and $Q_B$ are the 
canonical
partitions functions of
systems $A$ and $B$. To calculate $\Delta F$ using Jarzynski's
identity we introduce a parameter dependent Hamiltonian ${\cal H}(x,
\lambda)$ defined such that 
${\cal H}_A$ and ${\cal
H}_B$ are obtained for particular values of the control parameter,
${\cal H}(x, \lambda_A)={\cal H}_A(x)$ and ${\cal H}(x,
\lambda_B)={\cal H}_B(x)$. By switching the control parameter
$\lambda$ from $\lambda_A$ to $\lambda_B$ we can continuously
transform ${\cal H}_A(x)$ into ${\cal H}_B(x)$. If this is done
over a time $\tau$,
while the system evolves 
from particular
initial conditions $x_0$, the work performed on the system is
\begin{equation}
W=\int_0^\tau dt\; \left(\frac{\partial {\cal H}}{\partial
\lambda}\right) \dot \lambda .
\end{equation} 
Its value
depends on the initial conditions $x_0$ and on the
particular way the control parameter $\lambda(t)$ is switched from its
initial to its final value. According to Jarzynski's identity, the free
energy difference can be evaluated by averaging the work exponential
$e^{-\beta W}$
over many such transformations.  Specifically, this average is performed
over a canonical distribution of initial conditions in the initial
equilibrium state,
\begin{equation}
\exp(-\beta \Delta F)=\int dx_0\; \rho(x_0) \exp\{-\beta W(x_0)\},
\end{equation}
where $\rho(x_0)=\exp\{-\beta {\cal H}_A(x)\} / Q_A$. 

In fast switching simulations based on the Jarzynski identity,
non-equilibrium 
trajectories are generated by approximately integrating the equation
of motion, typically through a truncated Taylor expansion of the
time-evolving phase space point $x(t)$.  
The fidelity of
trajectories obtained in this way to true microscopic dynamics
is determined by the time interval over which
a low-order Taylor expansion is assumed to be accurate.
Usually, the time step is chosen to be small, so that the
total energy is nearly conserved when control parameters are held 
constant (in an isolated system)
\cite{ALLEN,FRENKEL_SMIT}. In this paper we show that fast switching trajectories
integrated with large time steps, while perhaps poor simulations
of dynamics, suffice to compute {\em exact} free energy differences.
This new approach, which can increase the efficiency of fast switching
simulations by up to two orders of magnitude, is based on a
generalization of Jarzynski's identity for general phase space
mappings\cite{jarTPS,targeted}.  Jarzynski's original expression
corresponds to the particular phase space mapping provided by the
dynamical propagator.
His result
is valid, however, for any
invertible phase space mapping.
One could just as well use a
concatenation of highly approximate molecular dynamics steps,
the result of integrating equations of motion to low order over large
time intervals, to map points in phase space.
Although such large time step trajectories are
not accurate dynamical pathways, expressions for the free
energy remain exact. Due to the reduced cost of large time step
trajectories, a considerable efficiency increase is possible.

The remainder of the paper is organized as follows. The formalism and
the justification of the large time step approach are presented in
section \ref{sec:formalism}. The efficiency of the resulting algorithm is
discussed in section \ref{sec:efficiency}. In section
\ref{sec:numerical} we demonstrate the validity of this algorithm by
calculating the 
reversible work to transform a
simple one-dimensional energy landscape, and that
to drag a particle through a Lennard-Jones fluid.  Conclusions are
given in section \ref{sec:conclusion}.

\section{Formalism}
\label{sec:formalism}

\subsection{Jarzynski's identity for phase space mappings}

The deterministic time evolution of a classical
many-particle system can be viewed as mapping every point in
phase space to another:
a system initially at $x_0$ will be located at $x_t=\phi_t(x_0)$ after
a time $t$. The function $\phi_t$ is called the {\em propagator} of the
system. Since the system evolves deterministically the point $x_t$ is
completely determined by the initial conditions $x_0$. The time
reversibility of equations of motion further ensures that such a
mapping is invertible, i.e., that from $x_t$ the corresponding
starting point $x_0$ can be uniquely determined, $x_0=\phi_{-t}(x_t)$.

Consider now a general invertible and differentiable mapping
\begin{equation}
x' = \phi(x).
\label{equ:map}
\end{equation}
that maps phase space point $x$ into phase space point $x'$. Here, the
mapping $\phi(x)$ takes the place of the propagator $\phi_t(x)$. For
such mappings Jarzynski has derived an expression akin to the
non-equilibrium work theorem \cite{targeted}. To introduce the
necessary notation we rederive this result. For this purpose we
consider the definition of the free energy difference,
\begin{equation}
\exp(-\beta \Delta F)=\frac{Q_B}{Q_A}=\frac{\int dx' \exp\{-\beta
{\cal H}(x', \lambda_B)\}}{Q_A}.
\end{equation}
Multiplying and dividing the integrand in the above equation with
$\exp\{-\beta {\cal H}(\phi^{-1}(x'), \lambda_A)\}$ we obtain
\begin{eqnarray}
\exp(-\beta \Delta F) & = & \int dx' \frac{\exp \{-\beta {\cal
 H}(\phi^{-1}(x'), \lambda_A)\}}{Q_A} \times \nonumber \\ & &
 \hspace{-1cm}\exp\{-\beta [{\cal H}(x', \lambda_B)-{\cal
 H}(\phi^{-1}(x'), \lambda_A)]\}.
\end{eqnarray}
A change of integration variables from $x'$ to $x=\phi^{-1}(x')$ yields
\begin{eqnarray}
\exp(-\beta \Delta F) & = & \int dx \left|{\partial \phi(x) \over
 \partial x} \right| \frac{\exp \{-\beta {\cal H}(x,
 \lambda_A)\}}{Q_A} \times \nonumber \\ & & \hspace{-1cm}\exp\{-\beta
 [{\cal H}(\phi(x), \lambda_B)-{\cal H}(x, \lambda_A)]\}.
\label{equ:dFW}
\end{eqnarray}
where $\left| \partial \phi(x) / \partial x \right|$ is the Jacobian
determinant of the mapping $\phi(x)$. The above equation suggests 
generalizing the
definition of work
\begin{equation}
W_\phi={\cal H}(\phi(x), \lambda_B)-{\cal H}(x, \lambda_A)-k_B T \ln
\left|{\partial \phi(x) \over \partial x} \right|.
\label{equ:Wphi}
\end{equation}
This ``work'' includes the energy change caused by switching the
control parameter from $\lambda_A$ to $\lambda_B$.  In addition
$W_\phi$ includes a term involving the Jacobian of $\phi(x)$, which
can be viewed as the work necessary to compress or expand the phase
space volume when applying the mapping $\phi(x)$. This entropic
contribution can be interpreted as ``heat'' absorbed during
the mapping. In fact, if we choose the mapping to be the system's
propagator, this term is exactly the heat. In this interpretation,
Equ.~(\ref{equ:Wphi}) is nothing other than an expression of the first
law of thermodynamics.

Using the work definition (\ref{equ:Wphi}) we can rewrite
Equ. (\ref{equ:dFW}) as
\begin{equation}
\exp(-\beta \Delta F) = \int dx \frac{\exp \{-\beta {\cal H}(x,
\lambda_A)\}}{Q_A} \exp\{-\beta W_\phi (x)\}
\end{equation}
or, as an average over the 
initial equilibrium distribution,
\begin{equation}
\exp(-\beta \Delta F)  =  \langle \exp\{-\beta W_\phi (x)\}\rangle.
\label{equ:jarphi}
\end{equation}
This equation can be viewed as a generalization of Jarzynski's
identity. If the mapping is chosen to be the propagator $\phi_t(x)$ of 
Newtonian dynamics,
the work $W_\phi$ equals the physical work $W$ carried out on the
system as it evolves from $x_0$ to $x_t$,
\begin{equation}
W= {\cal H}(x_t,\lambda_B)-{\cal H}(x_0,\lambda_A).
\end{equation}
and Equ.~(\ref{equ:jarphi}) reduces to Jarzynski's identity. In
deriving this result we have exploited the fact that Newtonian
dynamics conserves phase space volume -- even when a control parameter
changes with time, the Jacobian appearing in the definition
of $W_\phi$ [Equ. (\ref{equ:Wphi})] is unity.

\subsection{Long time step trajectories}

Instead of the propagator $\phi_t$, we can choose a sequence of
molecular dynamics steps as our mapping. 
Each of these steps, which are
designed to approximate the time evolution of the system over a small
time interval $\Delta t$, maps a phase point $x_i$ into a phase point
$x_{i+1}$. Equation~(\ref{equ:jarphi}) can be applied to a map defined by
$n$ such steps, together taking the initial point $x_0$ into a final
point $x_n=\phi_n(x_0)$. The expression for the work $W_\phi$ is
particularly simple for integrators such as the Verlet algorithm,
which both conserve phase space volume and are
time-reversible\cite{ALLEN,FRENKEL_SMIT}. In this case the Jacobian of the mapping
is unity, and, according to Equ.~(\ref{equ:Wphi}),
\begin{equation}
W_\phi(x_0)={\cal H}(x_n,\lambda_B) - {\cal H}(x_0,\lambda_A),
\end{equation}
so that
\begin{equation}
e^{-\beta \Delta F} = \langle \exp\{-\beta [{\cal
H}(x_n,\lambda_B)-{\cal H}(x_0,\lambda_A)]\} \rangle.
\label{equ:finite}
\end{equation}
This relation is exact regardless of the size of the time step $\Delta t$
used in applying these integrators. 

Equation~(\ref{equ:finite}) suggests the following algorithm. (1) A
canonical distribution $\rho(x_0)$ of initial conditions is sampled
with a Monte Carlo procedure or with an appropriately thermostatted
molecular dynamics simulation. (Note that in the latter case a
sufficiently small time step must be used in order to preserve the
correct equilibrium distribution.) (2) These initial conditions are then
used as starting points for 
fast switching
trajectories obtained by repeated application of the Verlet
algorithm. (3) During the integration the control parameter is changed
from $\lambda_A$ to $\lambda_B$. Since Equ.~(\ref{equ:finite}) is
exact for any size of the time step $\Delta t$, the chosen integration
time step can be
arbitrarily large, provided that the variables specifying the state of
the system (for instance, positions and momenta of all particles)
retain values that do not exceed the range a computer can
represent. We call this limit the {\em stability limit}. For each
trajectory the energy difference $W = {\cal H}(x_n,\lambda_B)-{\cal
H}(x_0,\lambda_A)$ is determined and used to calculate the exponential
average appearing in Equ.~(\ref{equ:finite}). Since large time step
trajectories are computationally less expensive, this algorithm
holds promise to increase the efficiency of fast-switching free energy
calculations. Whether this is actually the case depends on how the
work distribution is modified by the increase in time step length. In
Sec.~III we describe how to analyze the efficiency of
fast switching simulations with large time steps.

\subsection{Stochastic dynamics}

Often the dynamics of model molecular systems evolve by stochastic
equations of motion. Common examples include the Langevin equation
\cite{ZWANZIG},
\begin{eqnarray}
\dot{q} &=& p/m
\nonumber
\\
\dot{p} &=&  - {\partial {\cal H}\over \partial q} -\gamma p + \eta(t),
\label{equ:langevin}
\end{eqnarray}
where $\gamma$ is a friction coefficient and $\eta(t)$ is a
fluctuating random force; and deterministic dynamics coupled to
stochastic thermostats, such as the Andersen thermostat
\cite{ANDERSEN}. It has been shown that the Jarzinsky relation remains
valid also in these cases \cite{gavin_jstatphys,JARZ_PRE_97}. In this section we discuss 
the question whether the large time step approach discussed in the
previous section can be applied to stochastic dynamics as well.

We describe the stochastic component of these dynamics through a
``noise history'' $\eta(t)$.  In the case of Langevin dynamics the
noise history is, as the notation suggests, the trajectory of the
random force.
For a given realization of $\eta(t)$, the time
evolution of a stochastic system can be regarded as deterministic, and
we may write
\begin{equation}
x_t=\phi[x_0;\eta(t)],
\end{equation}
where the second argument in the deterministic map indicates its
dependence on the noise history $\eta(t)$. Since this mapping is
invertible and differentiable, 
Equ. (\ref{equ:jarphi}) applies for any
particular noise history,
\begin{equation}
\exp(-\beta \Delta F) = \int dx \rho(x) \exp\{-\beta W_\phi [x;
\eta(t)]\}\rangle,
\end{equation}
where we have averaged over canonically distributed initial conditions and 
\begin{eqnarray}
\lefteqn{\hspace{-0.6cm}W_\phi[x; \eta(t)]=}\hspace{0.5cm} \nonumber
\\ & & \hspace{-1.4cm}{\cal H}(\phi[x; \eta(t)], \lambda_B)-{\cal
H}(x, \lambda_A)-k_B T \ln \left|{\partial \phi[x;\eta(t)] \over
\partial x} \right|.
\end{eqnarray}
This result above is valid for any noise-dependent map
$\phi[x_0;\eta(t)]$, so we are free to choose a mapping comprised of
repeated application of Brownian dynamics steps \cite{ALLEN} with
arbitrary step size.  Because the result of averaging $\exp\{-\beta
W_\phi [x; \eta(t)]\}$ over initial conditions is completely
independent of $\eta(t)$, the remaining average over noise histories
is trivial, yielding
\begin{eqnarray}
\lefteqn{\exp(-\beta \Delta F) =} \hspace{0.8cm}\nonumber \\ & &
\hspace{-1.0cm}\int {\cal D}\eta(t) \int dx P[\eta(t)]\rho(x)
\exp\{-\beta W_\phi [x; \eta(t)]\}.
\label{equ:stochastic}
\end{eqnarray}
Here, the notation $\int {\cal D}\eta(t)$ indicates summation over all
noise histories and $P[\eta(t)]$ is the probability distribution for
observing a particular realization.

Interestingly, the above derivation implies that
Equ. (\ref{equ:stochastic}) can be applied to mappings with a
completely arbitrary stochastic component. In particular, it is not
necessary that the magnitude of stochastic fluctuations be related in
any way to the rate of dissipation.  In order to preserve a canonical
distribution, the Langevin equation must be supplemented with such a
constraint on the statistics of $\eta(t)$ (henceforth assumed to 
be Gaussian white noise):
\begin{equation}
\langle \eta(t) \eta(t')\rangle = 2 k_{\rm B}T \gamma \delta(t-t').
\label{equ:flucdiss}
\end{equation}
This fluctuation-dissipation relation, ensuring detailed balance, is a
necessary condition for the applicability of Jarzynski's original
identity concerning the exponential average of work defined in the
conventional way.  Obtaining that result from
Equ.~\ref{equ:stochastic} is not nearly as straightforward as was the
analogous task for deterministic dynamics.  For stochastic mappings
the Jacobian determinant does not directly correspond to heat, even in
the limit of small time step size.  Instead, for a trajectory of
length $\tau$
\begin{equation}
\lim_{\Delta t \rightarrow 0}\left|{\partial \phi[x;\eta(t)] \over
\partial x} \right| = e^{- n_{\rm f}\gamma \tau},
\label{equ:lang_jac}
\end{equation}
where $n_{\rm f}$ is the number of momentum degrees of freedom.  The
volume of a phase space element evolving under Langevin dynamics thus
decays steadily as time evolves and has no contribution from the
fluctuating random force.  Mathematically, this phase space
compression arises from the systematic damping of kinetic energy
through the friction term in Equ.~\ref{equ:langevin}.  Physically, its
cancellation of contributions from heat ${\cal Q}$ in the exponential
average of Equ.~\ref{equ:stochastic} is a subtle consequence of
detailed balance.  Crooks has shown that, for trajectories generated
by any balanced dynamical rules, $e^{-\beta {\cal Q}}$ is equivalent
to the ratio of probability densities of forward and time-reversed
pathways.  This ratio is closely related to a mapping's Jacobian, as
will be shown in Sec. \ref{sec:discussion}.

The practical utility of our large time step result for free energy
differences is compromised in the specific case of Langevin dynamics
by at least two issues.  First, the rapid decay of $\left|{\partial
\phi[x;\eta(t)] / \partial x} \right|$ could damp all but the
largest fluctuations, making convergence of the average in
Equ.~\ref{equ:stochastic} problematic.  Second, an exact
calculation of the Jacobian for large time steps is cumbersome when
many degrees of freedom interact.  The insensitivity
Equ. (\ref{equ:stochastic}) to the form of stochastic noise might be
exploited to offset these problems, but it is not obvious how to do
so.

\section{Efficiency analysis}
\label{sec:efficiency}

In a straightforward application of the fast switching procedure, the
free energy difference $\Delta F$ is estimated from a finite sample of
$N$ trajectories originating from canonically distributed initial
conditions. Here we assume that these initial conditions are
statistically uncorrelated samples.
Defining
\begin{equation}
X\equiv\exp(-\beta W_\phi),
\end{equation}
we can write the free energy difference for finite $N$ as
\begin{equation}
\label{equ:dF_finite}
\Delta \overline{F}_N \equiv -k_{\rm B}T\ln \overline{X}_N.
\end{equation} 
where $ \overline{X}_N$ is the average of $X$ over $N$ independent
trajectories:
\begin{equation}
\overline{X}_N\equiv\frac{1}{N}\sum_{i=1}^{N} X^{(i)}.
\end{equation}
Here, $X^{(i)}$ is the value of $X$ obtained from the $i$-th
trajectory.  We now repeat this entire procedure $M$ times (generating
a total of $MN$ trajectories) and average over all $M$ resulting free
energy estimates $\Delta\overline{F}_N^{(j)}$. The limiting result of this
protocol is
\begin{equation}
\langle \Delta \overline{F}_N \rangle\equiv \lim_{M\rightarrow
\infty}\frac{1}{M}\sum_{j=1}^{M}\Delta \overline{F}_N^{(j)}.
\end{equation}
Due to the non-linearity of the logarithm relating the average $
\overline{X}_N$ to the free energy estimate $\overline{F}_N$ (see
Equ. (\ref{equ:dF_finite})), $\langle \Delta \overline{F}_N \rangle$
differs from the true free energy difference $\Delta F$ even in the
limit of infinitely many repetitions. For sufficiently large $N$ this
deviation, or bias, is given by
\cite{BUSTAMANTE,ZUCKERMAN_WOOLF,jarTPS}
\begin{equation}
b_N\equiv\langle \Delta \overline{F}_N \rangle-\Delta F=\frac{k_{\rm
B}T}{2N}\frac{\langle (\delta X)^2 \rangle}{\langle X\rangle^2},
\end{equation}
where the fluctuation $\delta X\equiv X-\langle X\rangle$ is the
deviation of $X$ from its average value. This equation is obtained by
expanding the logarithm in Equ. (\ref{equ:dF_finite}) in powers of
relative fluctuations of $\overline X_N$ and truncating the expansion
after the quadratic term.

In addition to the systematic bias, we must account for random
error in each free energy estimate $\overline F_N^{(j)}$.
Assuming the statistical errors of different samples to be
uncorrelated, and denoting their variance as
\begin{equation}
\sigma_N^2 \equiv \langle [\Delta \overline{F}_N- \langle \Delta
\overline{F}_N \rangle]^2\rangle,
\end{equation}
we obtain the total mean squared deviation from the true free energy
difference,
\begin{equation}
\epsilon_N^2 \equiv \langle [\Delta \overline{F}_N- \Delta
F]^2\rangle=b_N^2+\sigma_N^2.
\end{equation}
Note that the bias decays with sample size as $b_N \propto 1/N$ for
large $N$, while the scale of random error decays only as
$\sigma_N\propto 1/\sqrt{N}$.
Thus, for sufficiently large $N$ only the statistical errors in the
free energy estimate are relevant, and we can safely approximate
\begin{equation}
\epsilon_N^2 = \sigma_N^2= \frac{k_{\rm B}^2T^2}{N}\frac{\langle
(\delta X)^2 \rangle}{\langle X\rangle^2}.
\label{equ:epsilon}
\end{equation}

From Equ. (\ref{equ:epsilon}) one can determine the number of
trajectories $N_\epsilon$ necessary to obtain a certain level of error
$\epsilon$,
\begin{equation}
N_\epsilon= \frac{k_{\rm B}^2T^2}{\epsilon^2}\frac{\langle (\delta
X)^2 \rangle}{\langle X\rangle^2}.
\end{equation}
The computational cost of each trajectory is roughly proportional to
the number of required force calculations and hence proportional to
the number of steps $n=\tau /\Delta t$ necessary to generate a
trajectory of length $\tau$. Neglecting the cost of the generation of
initial conditions we can thus define a normalized computational cost
\begin{equation}
C_{\rm CPU}\equiv n \frac{\langle (\delta X)^2 \rangle}{\langle
X\rangle^2}=\frac{\tau}{\Delta t}\frac{\langle (\delta X)^2
\rangle}{\langle X\rangle^2}.
\label{equ:CPU}
\end{equation}
This computational cost $C_{\rm CPU}$ is the CPU-time required to
obtain an accuracy in the free energy of $\epsilon=k_{\rm B}T$
measured in units of the CPU-time required to carry out one single
molecular dynamics time step. Note that while $\langle X\rangle$ is 
independent from the stepsize $\Delta t$, the mean square 
fluctuations $\langle (\delta X)^2\rangle$ depend on it. To
determine an optimal time step size for a fast-switching free energy
calculation, we must minimize the entire quantity $C_{\rm CPU}$.
In the following
section we will present calculations of this normalized CPU-time as a
function of the stepsize $\Delta t$ for two different models.

\section{Numerical results}
\label{sec:numerical}

\subsection{One-dimensional system}

We first study the effect of large time step integration for a simple
model introduced by Sun \cite{SUN}. In this one-dimensional model a
point particle of unit mass moves in a potential that depends on a
control parameter $\lambda$. The Hamiltonian for the system as a
function of the position $q$ and the momentum $p$ of the moving
particle is
\begin{equation}
\label{sunham}
\mathcal{H}( q, p; \lambda) = \frac{1}{2}p^2 + q^4 - 16(1 - \lambda)q^2 ,
\end{equation}  
Here, all quantities are scaled to be unitless. For $\lambda=0$ the
potential has two symmetric minima located at $q=\pm\sqrt{8}$
separated by a barrier of height $\Delta E=64$ located at $q=0$. For
$\lambda=1$ the potential energy function reduces to
a single quartic well.
For a given
value of the control parameter $\lambda$ the partition function is
$\int dq dp \exp\{-\beta \mathcal{H}(q,p,\lambda)\}$. The free energy
difference between the two states corresponding to $\lambda=1$ and
$\lambda=0$, respectively, can be calculated analytically to be
$62.9407 k_{\rm B}T$ \cite{jarTPS}.

For this model we carried out fast switching simulations using
different time steps $\Delta t$ ranging from $\Delta t=0.002$ to
$\Delta t=0.1$. The equations of motion were integrated with the
velocity Verlet algorithm \cite{ALLEN,FRENKEL_SMIT}, yielding positions
$q_t$ and momenta $p_t$ as a function of time $t$. For this model the
Verlet algorithm becomes unstable for time steps larger than $\Delta t
= 0.1$. In all cases the total trajectory length was $\tau=10$,
corresponding to a transformation sufficiently 
gradual to allow accurate calculation of the free energy difference
and of the mean square fluctuations $\langle (\delta X)^2\rangle$. For
each time step size we integrated $10^6$ trajectories
from initial conditions sampled from a canonical distribution
(with $\lambda=0$) using a Monte Carlo procedure.
Along the trajectories the control parameter was varied
from its initial to its final value. 
More precisely, after
each velocity Verlet step, carried out at constant $\lambda$, the
control parameter was increased by $1/n$ where $n=\tau / \Delta t$ is
the number of time steps in the trajectory. At the end of each
trajectory the work $W=\mathcal{H}(q_\tau, p_\tau;1)-\mathcal{H}(q_0,
p_0;0)$ was calculated and added to the exponential average. (Note that,
when advancing time in large steps,
it is important to use this generalized definition of work rather than 
summing estimates of the physical work
performed during each stepwise change of the control parameter.)

\begin{figure}[h]
\centerline{\includegraphics[width=7.0cm]{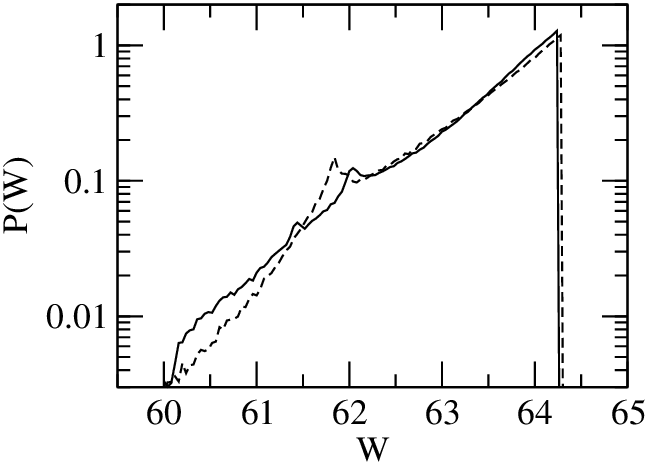}}
\caption{Work distributions $P(W)$ obtained for the Sun model for a
trajectory length $\tau=10$ and step sizes $\Delta t=0.1$ (solid line)
and 0.01 (dashed line). Work distributions for all time steps smaller
than $\Delta t=0.01$ are indistinguishable from the distribution for
$\Delta t=0.01$ on the scale of the figure.}
\label{fig:PW_sun}
\end{figure}

\begin{figure}[h]
\centerline{\includegraphics[width=7.0cm]{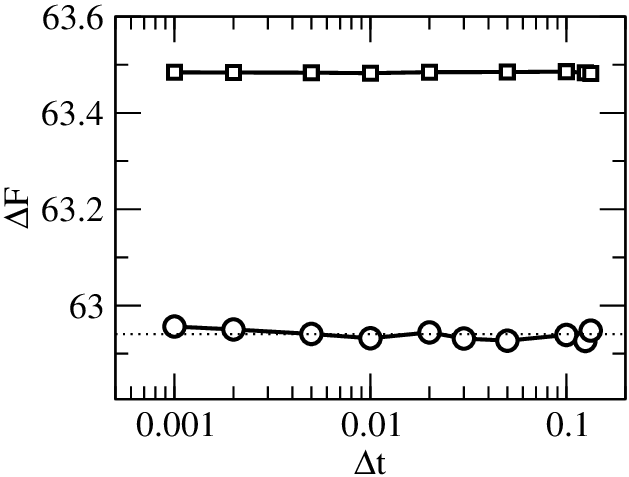}}
\caption{Free energy differences (circles) obtained for the Sun model
from fast switching trajectories of length $\tau=10$ with different
step sizes $\Delta t$. The dotted line denotes the exact free energy
difference. Also shown is the average work $\langle W \rangle$
(squares).}
\label{fig:dF_sun}
\end{figure}

Work distributions $P(W)$ obtained for different step sizes $\Delta t$
(and hence for different numbers of steps per trajectory) are shown in
Fig. \ref{fig:PW_sun}.  $P(W)$ deviates visibly from its
small time step limit
only for the largest step size, $\Delta t=0.1$ These differences
originate in the inaccuracy of the integration algorithm for large
step sizes. Even though the work distribution varies with the size of
integration steps, the resulting free energies show no step size
dependence (see Fig. \ref{fig:dF_sun}), provided the stability limit
of the integration algorithm is not exceeded.

\begin{figure}[h]
\centerline{\includegraphics[width=7.0cm]{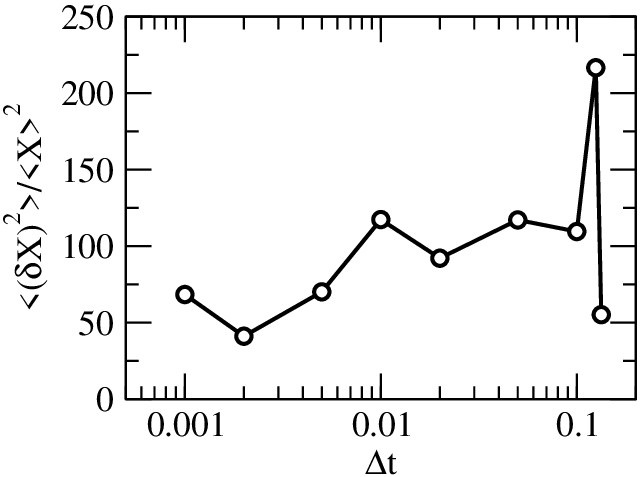}}
\caption{Relative fluctuations $\langle (\delta X)^2\rangle / \langle
X \rangle^2$ for the Sun model as a function of the step size $\Delta
t$.}
\label{fig:fluc_sun}
\end{figure}

To quantify the statistical error in the free energy estimates shown in
Fig.~\ref{fig:dF_sun} we have calculated the relative fluctuations
$\langle (\delta X)^2 \rangle / \langle X \rangle^2$, which according
to Equ.~(\ref{equ:epsilon}) determine the mean squared error
$\epsilon_N^2$. These relative fluctuations are plotted in
Fig.~\ref{fig:fluc_sun}. The irregular shape of this curve reflects
changes in 
the features (such as the peak near $W=62$) of the work distributions
shown in Fig. \ref{fig:PW_sun}.

\begin{figure}[h]
\centerline{\includegraphics[width=7.0cm]{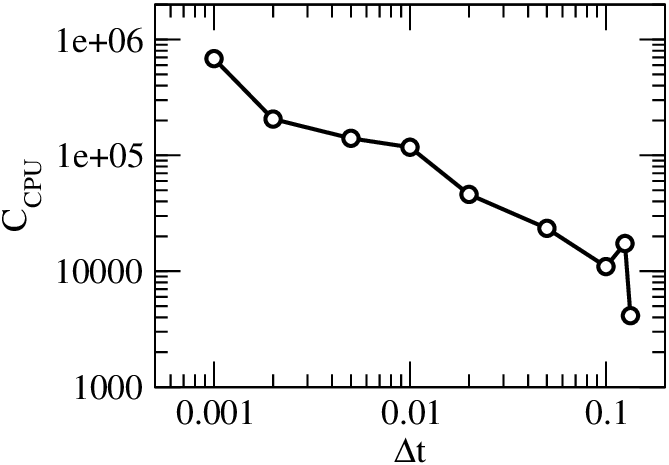}}
\caption{Normalized CPU time $C_{\rm CPU}$ for the Sun model as a
function of the step size $\Delta t$ for $k_{\rm B}T=1$. For all step sizes 
the total trajectory length was $\tau=1$. This result indicates that for 
this model the  computational cost of a free energy calculation with a 
given error decreases for increasing step size until the stability
limit is reached.}
\label{fig:CPU_sun}
\end{figure}

The computational cost $C_{\rm CPU}$ follows from the relative
fluctuations and is shown as a function of the step size in
Fig. \ref{fig:CPU_sun}. Over the range of time steps depicted in the
figure the computational cost decreases from about $10^6$ to about
$10^4$. Thus, the fast switching simulation can be accelerated by two
orders of magnitude if the conservative step size of $\Delta
t=0.001$ is replaced by a step size $\Delta t=0.1$ near the stability
limit. Although in the latter case the equations of motion are not
faithfully solved (see Sec. \ref{sec:discussion}, Fig. \ref{fig:Wepsilon}), 
the expression for the free energy remains exact.

\subsection{Dragged particle in Lennard-Jones fluid}

Our second example involves a system with many degrees of freedom and
is therefore likely to be more relevant for typical molecular systems
of interest.  Spepcifically, we tested the large-timestep version of
the Jarzynski identity for a particle dragged through a Lennard-Jones
fluid. In these simulations, a tagged particle is coupled to a
harmonic trap 
whose minimum
is shifted from one position to another while the
system evolves in time. If this process is carried out at a finite
rate, work is performed on the system by the moving
trap. Nevertheless, the free energy difference between the two states
corresponding to the initial and final position of the trap vanishes
due to symmetry.

The time-dependent Hamiltonian for our $M$-particle system is
\begin{equation}
\label{LJHamiltonian}
\mathcal{H}(p,q,t) = \sum_{i=1}^M \frac{{\bf p}_i^2}{2m_i} +
\sum_{i<j}^M v(r_{ij}) + \frac{k}{2} ({\bf r}_1- {\bf R}(t))^2 .
\end{equation}
where ${\bf p}_i$ is the momentum of particle $i$, ${\bf r}_i$ is the
position of particle $i$, $r_{ij}$ is the distance between particles
$i$ and $j$ and the second sum on the right hand side extends over all
particle pairs. The particles interact via the Lennard-Jones potential
\begin{equation}\label{LJPotential}
v(r) = 4 \varepsilon \left[\left(\frac{\sigma}{r}\right)^{12} -
\left(\frac{\sigma}{r}\right)^{6} \right].
\end{equation}
Here, $\varepsilon$ and $\sigma$ are parameters describing the depth of
the potential well and the interaction range of the strongly repulsive
core, respectively.  The last term in Equ. (\ref{LJHamiltonian})
describes the potential energy of one particular particle (we
arbitrarily pick particle $1$) in a harmonic trap with force constant
$k$. In Equ. (\ref{LJHamiltonian}) ${\bf R}(t)$ denotes the
time-dependent position of the trap's minimum,
which is moved from the origin in the $x$-direction with constant speed $\nu$,
\begin{equation}
\label{equ:trap}
{\bf R}(t)={\bf e}_x\nu t,
\end{equation}
where ${\bf e}_x$ is the unit vector in $x$-direction. During a total
time $\tau$ the trap is displaced by an amount $L=\nu \tau$. While the
trap's minimum roughly determines the position of
particle $1$, this particle does fluctuate about ${\bf R}(t)$.
The work $W$ performed on the system along a particular trajectory
from $\{q_0, p_0\}$ to $\{q_\tau, p_\tau\}$ is given by
\begin{equation}
W_\phi={\cal H}(q_\tau,p_\tau;\tau)-{\cal H}(q_0,p_0;0).
\end{equation}
Since the free energy of the system does not depend on the
trap's location, $\Delta F=0$,
the exponential work average carried out over many
realizations of this process is
\begin{equation}
\label{LJJarzynski}
\langle \exp(-\beta W_\phi ) \rangle = \exp(-\beta \Delta F)=1.
\end{equation}
Here, angular brackets indicate a canonical average for a fixed
position of the trap.

We have carried out a fast switching procedure for $M = 108$ particles
of unit mass in a three-dimensional, cubic simulation box with
periodic boundary conditions. The fluid's density, $\rho = 0.8 \sigma
^{-3}$, is roughly that at the triple point, and its initial temperature
places it in the liquid phase. Canonically distributed
initial conditions were generated by a molecular dynamics simulation,
employing an Andersen thermostat at temperature $k_{\rm B} T
/ \varepsilon = 1.0$ \cite{ANDERSEN}, with the trap fixed at the origin. 
For the generation of initial conditions, the equations of motions
were integrated with the velocity Verlet algorithm and a time step of
$dt = 0.001 \sigma (\varepsilon/m)^{1/2}$.  The state of the system was
recorded every 50 steps along this equilibrium trajectory, providing
an ensemble of intial conditions for the fast switching procedure.

\begin{figure}[h]
\centerline{\includegraphics[width=7.0cm]{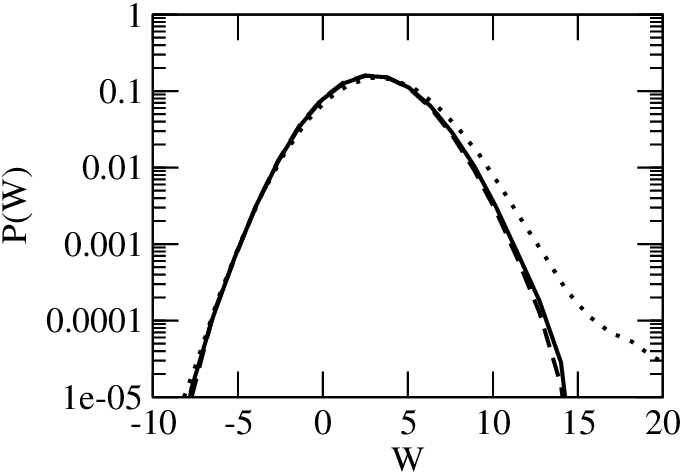}}
\caption{Work distributions $P(W)$ for a particle dragged through a
Lennard-Jones fluid for step size $\Delta t = 0.001\, \sigma (\varepsilon
/m)^{1/2}$ (solid line), step size $\Delta t = 0.015\, \sigma
(\varepsilon /m)^{1/2}$ (dashed line) and step size $\Delta t = 0.02\,
\sigma (\varepsilon /m)^{1/2}$ (dotted line).}
\label{fig:PW_GCM}
\end{figure}

\begin{figure}[h]
\centerline{\includegraphics[width=7.0cm]{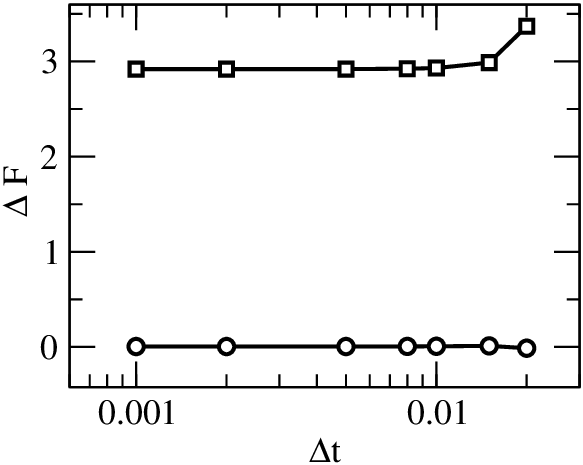}}
\caption{Free energy differences (circles) for a particle dragged
through a Lennard-Jones fluid as a function of step size $\Delta t$. The 
trajectory length was $\tau=1.2 \sigma (\varepsilon /m)^{1/2}$ for all
step sizes. Also shown is the average work
$\langle W \rangle$ (squares).}
\label{fig:dF_GCM}
\end{figure}

From these initial conditions we generated fast switching
trajectories of total length $\tau = 1.2 \sigma (\varepsilon/m)^{1/2}$
with different time steps ranging from $\Delta t = 0.001 \sigma
(\varepsilon/m)^{1/2}$ to $\Delta t = 0.02 \sigma (\varepsilon/m)^{1/2}$.
The corresponding pathways ranged in number of steps 
from $n=1200$ to $n=60$, respectively. In each case the total
displacement of the trap from its initial to its final position was
$L=0.5\sigma$ corresponding to a velocity of $\nu=5/12
(m/\varepsilon)^{1/2}$. The velocity Verlet algorithm without thermostat
was used to integrate Newton's equations of motion. Along these
trajectories the particle trap was displaced stepwise by a small
distance $L/n$ after each molecular dynamics step. Work distributions
obtained in this manner are shown in
Fig. \ref{fig:PW_GCM} for three different step sizes.
Free energy differences $\Delta F$ and the average work $W$ calculated
in these simulations are depicted in Fig. \ref{fig:dF_GCM}.

\begin{figure}[h]
\centerline{\includegraphics[width=7.0cm]{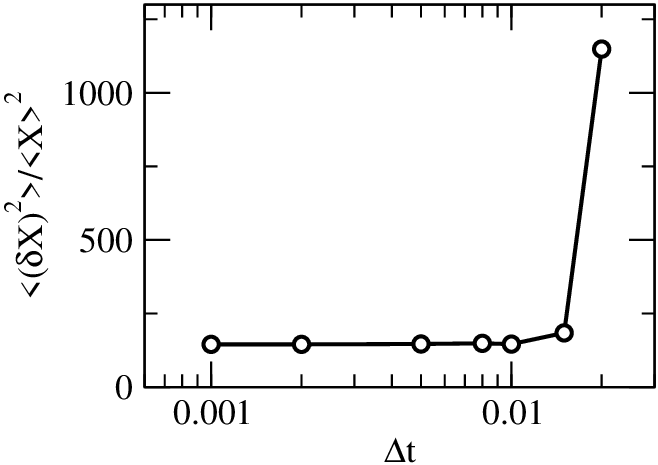}}
\caption{Relative fluctuations $\langle (\delta X)^2\rangle / \langle
X \rangle^2$ for the particle dragged through a Lennard-Jones fluid as
a function of the step size $\Delta t$.}
\label{fig:fluc_GCM}
\end{figure}

\begin{figure}[h]
\centerline{\includegraphics[width=7.0cm]{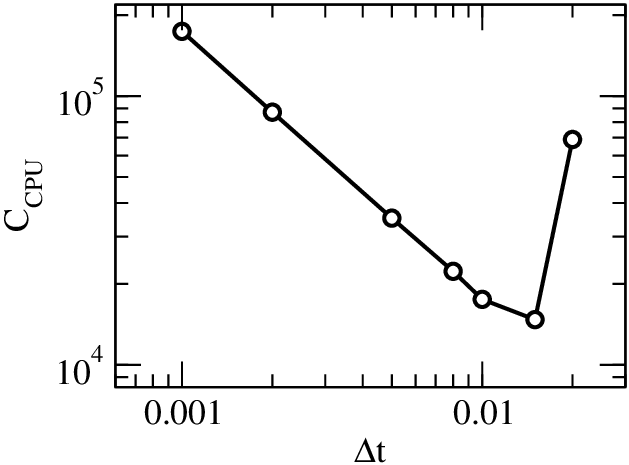}}
\caption{Normalized CPU time $C_{\rm CPU}$ for the particle dragged
through a Lennard-Jones fluid of $M=108$ particles at a density 
$\rho=0.8\sigma^{-3}$ as a function of the  step size $\Delta t$ at 
$k_{\rm B}T/\varepsilon=1$. The particle is dragged through
the fluid by a parabolic potential with force constant 
$k=10^3(\varepsilon / \sigma^2)$ moving at 
constant speed $\nu=5/12 (m/\varepsilon)^{1/2}$. For all step sizes the total 
trajectory length  was $\tau = 1.2 \sigma (\varepsilon/m)^{1/2}$. 
Also for this model the cost of a free energy calculation decreases
up to step sizes just short of the stability limit.}
\label{fig:CPU_GCM}
\end{figure}

We used the relative fluctuations $\langle (\delta X)^2\rangle /
\langle X \rangle^2$ (see Fig. \ref{fig:fluc_GCM}) to calculate the
normalized CPU time $C_{\rm CPU}$, which is plotted in
Fig. \ref{fig:CPU_GCM}.  The computational effort required to obtain a
specific accuracy decreases with increasing step size until the
largest value, $\Delta t = 0.02 \sigma (\varepsilon/m)^{1/2}$. Just as in
the schematic one-dimensional example, the most efficient
fast-switching calculation for dragging a particle through a dense
fluid is obtained for step sizes close to the stability limit.


\section{Discussion}
\label{sec:discussion}

Molecular dynamics simulations carried out with large time steps do
not faithfully reproduce the dynamics of any system with
configuration-dependent forces. If such large time steps are used in
the generation of fast switching trajectories to calculate free energy
differences on the basis of Jarzynski's identity, integration errors
lead to work distributions differing from those obtained in the small
time step limit. Nevertheless, as we have shown in
Sec. \ref{sec:formalism}, the Jarzynski identity remains exactly valid
in principle for time steps of arbitrary size.  As a practical matter
the stability limit provides an upper bound to the step size.  Since
the computational cost of molecular dynamics trajectories is
proportional to the number of integration steps, large time steps can
be beneficial. Whether an increase of computational efficieny can
really be achieved depends on how the relative squared fluctuations
$\langle (\delta X)^2 \rangle / \langle X \rangle^2$ scale with the
size of the integration time step. If these fluctuations increase with
$\Delta t$ sublinearly, using large time steps is advantageous.

In both of the numerical examples presented in
Sec.~\ref{sec:numerical}, increasing the size of the time step well
beyond the range appropriate for equilibrium simulations proved
favorable.  Resulting efficiency increases reached up to two orders of
magnitude. Here we ask more generally, and with complex molecular
systems in mind, what circumstances should allow
large time step integration to improve upon fast switching
simulations. 
To answer that question it is convenient to write the work $W$
performed on the system during the transformation as sum of two
parts. This separation is particularly natural if the control
parameter $\lambda$ is changed stepwise, i.e., if each step is
performed at constant $\lambda$ and is followed by an increase in the
control parameter by an increment $\Delta \lambda=1/n$. 

We define the first contribution to the generalized work $W_\phi$
as the change in energy due to the changes in control parameter for
fixed phase points,
\begin{equation}
W_{\lambda}\equiv \sum_{i=1}^{n}{\cal H}[x_i, i \Delta
\lambda]-{\cal H}[x_i, (i-1) \Delta \lambda].
\label{equ:Wham}
\end{equation}
Here, $x_i$ is the phase space point reached after $i$ integration
steps starting from phase space point $x_0$.  In Crooks's
considerations of systems out of equilibrium, $W_\lambda$ is precisely
the physical work exerted by the change of control parameter
\cite{gavin_jstatphys}.

In a system that evolves according to Newton's equations with a
time-independent potential energy function, the total energy is a
constant of the motion.
But when these equations are integrated approximately over finite
time steps, the energy of the system is not perfectly
conserved. Summing the energy changes due to integration error 
in the intervals between stepwise increases of $\lambda$, we
obtain the other contribution to $W_\phi$:
\begin{equation}
W_{\epsilon}\equiv \sum_{i=0}^{n-1}{\cal H}[x_{i+1}, i \Delta
\lambda]-{\cal H}[x_i, i \Delta \lambda].
\label{equ:Weps}
\end{equation}

To summarize our decomposition of $W_\phi$, the work $W_{\lambda}$
involves changes in control parameter at fixed phase space points,
while the less physical contribution $W_\epsilon$ involves changes in
the phase space point at constant control parameter. The total work
performed during the transformation is the sum of these two
quantities,
\begin{equation}
W=W_{\lambda}+W_{\epsilon}.
\end{equation}
In the limit of very small time steps, the energy of the system is
conserved whenever $\lambda$ is constant. In this case, the ``error''
work $W_\epsilon$ vanishes and the entire work is caused by changes in
the control parameter, $W= W_{\lambda}$.  If a system is driven away
from equilibrium by a stepwise increase of the control parameter, the
error work serves as a simple measure for the accuracy of approximate
numerical integration.

\begin{figure}[h]
\centerline{\includegraphics[width=7.0cm]{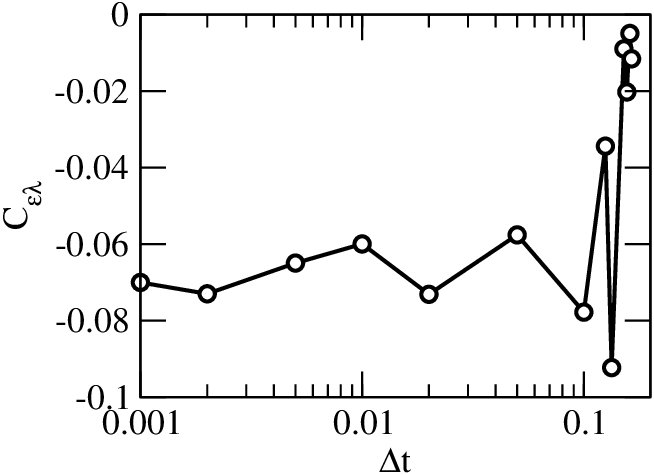}}
\caption{Correlations $C_{\epsilon\lambda}$ between control parameter
work and integration error work as a function of the step size for the
Sun model.}
\label{fig:corr}
\end{figure}

Neglecting correlations between $W_\lambda$ and $W_\epsilon$ for the
moment, we can write
\begin{equation}
\langle \exp(-\beta \Delta W) \rangle=\langle \exp(-\beta W_\lambda)
\rangle \langle \exp(-\beta W_\epsilon) \rangle.
\end{equation}
Accordingly, the free energy change is the sum of two terms
originating from $W_\lambda$ and $W_\epsilon$,
\begin{eqnarray}
\Delta F & = & \Delta F_\lambda + \Delta F_\epsilon \nonumber \\
&= & -k_{\rm B}T \ln \langle e^{-\beta W_\lambda} \rangle 
      -k_{\rm B}T \ln \langle e^{-\beta W_\epsilon} \rangle.
\end{eqnarray}
This separation of the free energy difference into two terms related
to $W_\lambda$ and $W_\epsilon$, respectively, is only strictly valid
if fluctuations in
\begin{equation}
X_\lambda\equiv \exp(-\beta W_\lambda) \quad \text{and} \quad
X_\epsilon \equiv \exp(-\beta W_\epsilon)
\end{equation}
are statistically independent.  This supposition cannot be assumed
a priori to be the case. For the two models treated numerically in
this study we have calculated the correlation
\begin{equation}
C_{\lambda\epsilon}\equiv \frac{\langle \delta X_\lambda \delta
X_\epsilon\rangle} {\sqrt{\langle (\delta X_\lambda)^2 \rangle \langle
(\delta X_\epsilon)^2\rangle}},
\end{equation}
where $ \delta X_\lambda \equiv X_\lambda - \langle X_\lambda\rangle$
and $ \delta X_\epsilon \equiv X_\epsilon - \langle
X_\epsilon\rangle$.  This coefficient quantifies correlations between
the two variables $X_\lambda$ and $X_\epsilon$. While in the absence
of correlations $C_{\lambda \epsilon}=0$, perfect correlation
(anticorrelation) leads to $C_{\lambda \epsilon}=1$ ($C_{\lambda
\epsilon}=-1$). Correlations computed numerically for the Sun model
are depicted in Fig. \ref{fig:corr} as a function of the time step
$\Delta t$. For all time step sizes $X_\lambda$ and $X_\epsilon$ are
only weakly (anti)correlated.  The assumption of statistical
independence is justified in these cases.

Under the same assumption
the statistical error of a fast switching free energy calculation
with large time steps can also be written as the sum of two distinct
contributions.  Absent correlations between $W_\lambda$ and
$W_\epsilon$, Equ. (\ref{equ:epsilon}) becomes
\begin{equation}
\epsilon_N^2 = \frac{k_{\rm B}^2T^2}{N}\left[\frac{\langle (\delta
X_\lambda)^2 \rangle}{\langle X_\lambda\rangle^2}+\frac{\langle
(\delta X_\epsilon)^2 \rangle}{\langle X_\epsilon\rangle^2}\right].
\label{equ:epsilon_sum}
\end{equation}
Thus, the total mean squared error is the sum of the mean squared
errors of the free energies related to the control parameter work and
the integration error work, respectively.  One potentially substantial
difference between these two error contributions is their dependence
on system size. Often, the control parameter $\lambda$ acts only on a
small subset of the system's degrees of freedom. In calculating the
chemical potential of an electrically neutral species by particle
insertion, for instance, the solute interacts only with a small number
of other particles near the insertion point. Similarly, the
transformation of a residue of a protein into another one mainly
affects only a local group of interactions. In such cases, the work
$W_\lambda$ resulting directly from 
changes in the control parameter quickly saturates with growing system
size. The work $W_\epsilon$, originating in the inaccuracy of the
integration algorithm, has a different system size dependence. Since
all degrees of freedom contribute to the integration error,
$W_\epsilon$ is expected to grow linearly with the system size. Thus,
for sufficiently large systems the second term on the right hand side
of Equ. (\ref{equ:epsilon_sum}) might become dominating for long time
steps, thus limiting the maximally possible efficiency gain. (It is
for similar reasons that hybrid Monte Carlo-molecular dynamics
simulations decline efficiency for large systems \cite{DUANE}.) But as
long as the integration error is small compared to work done by
changes in control parameter, using large time steps should remain
advantageous.

In our discussion of the efficiency of the large time step fast
switching approach we have so far neglected the computational cost
associated with generating initial conditions. Often, starting points
for the fast switching trajectories are generated with an
appropriately thermostatted molecular dynamics simulation. It is
important that these simulations are carried out with a time step of
conventionally small size.  Otherwise the distribution of initial
conditions can differ from the necessary canonical one. Thus, the
large time step approach can reduce only the computational cost
associated with generating the non-equilibrium trajectories.
This latter contribution is dominant by far in most cases.

The issues discussed in this paper have some interesting implications
for molecular dynamics simulations carried out at equilibrium as
well. If the control parameter $\lambda$ is not changed as the system
evolves in time, no work $W_\lambda$ is done on the system. In this
case the free energy difference $\Delta F$ also vanishes, since
initial and final Hamiltonians are identical. Nevertheless, due
to the imperfection of a finite time step integration algorithm, the
energy is not strictly conserved, and some ``error'' work $W_\epsilon$
is performed. According to Equ. (\ref{equ:finite}) the statistics of
such energy errors obeys
\begin{equation}
\langle \exp(-\beta W_\epsilon)\rangle=1.
\label{equ:Wepsilon}
\end{equation}
This result implies that the distribution of integration errors cannot
be symmetric 
about $W_\epsilon=0$. Rather, positive errors are in loose terms
more likely than negative ones. For a Gaussian error distribution,
Equ.~(\ref{equ:Wepsilon}) relates the average error to the width of
the error distribution,
\begin{equation}
\langle W_\epsilon \rangle = \frac{1}{2}\beta ^2 \sigma_\epsilon^2.
\end{equation}
For more complicated error distributions, this same result is obtained
from a cumulant expansion of Equ. (\ref{equ:Wepsilon}) truncated after
the second term. Thus, for an error distribution that is Gaussian
and/or sufficiently narrow, the average error is positive.  
This demonstration that the average energy of a simulated canonical
ensemble drifts upward with time makes no reference to the details of
molecular interactions.

\begin{figure}[h]
\centerline{\includegraphics[width=7.0cm]{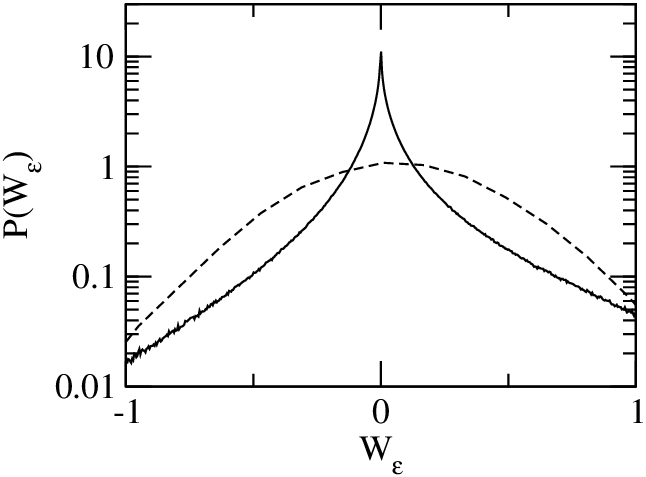}}
\caption{Distributions $P(W_\epsilon)$ of the integration error
$W_\epsilon$ at constant control parameter. Solid line: Sun model for
$\lambda=0$, $\tau=10$, and $\Delta t=4/30$; dashed line:
Lennard-Jones fluid with one particle in fixed trap, $\tau=1.2 \sigma
(\epsilon/m)^{1/2}$, $\Delta t=0.015\, \sigma (\epsilon /m)^{1/2}$.}
\label{fig:Wepsilon}
\end{figure}

We have verified Equ. (\ref{equ:Wepsilon}) both for the Sun model and
for the Lennard-Jones fluid. In both cases we have determined the
distribution $P(W_\epsilon)$ of the integration error $W_\epsilon$ for
a large integration step $\Delta t$ and constant $\lambda$. The
averages were carried out in the respective canonical ensembles
corresponding to $\lambda=0$. Two typical distributions are shown in
Fig. \ref{fig:Wepsilon}. For the Sun model (solid line) $P(W_\epsilon)$ 
is strongly non-Gaussian and asymmetric about
$W_\epsilon=0$, with positive errors more likely than negative
ones. While the average error of $\langle W_\epsilon \rangle=0.041$ is
clearly positive, the exponential average $\langle \exp(-\beta
W_\epsilon)\rangle$ is unity with high accuracy. For the Lennard-Jones
fluid with one particle in a fixed trap the distribution of
integration errors is approximately Gaussian, with an average of
$\langle W_\epsilon \rangle=0.064$. Also in this case the exponential
work average is unity.

Formulating changes in control parameter as a sequence of discrete
steps is also convenient for demonstrating the equivalence between
our identity for stochastic mappings,
Equ. (\ref{equ:stochastic}), and Jarzynski's
original identity.  During periods when $\lambda$ is fixed,
stochastic evolution subject to a suitable
fluctuation-dissipation relation, such as Equ. (\ref{equ:flucdiss}), satisfies detailed
balance in the small time step limit.  Specifically,
\begin{equation}
\rho_i(x) p(x \rightarrow x') =
\rho_i(x') \hat{p}({x}' \rightarrow {x})
\label{equ:balance}
\end{equation}
where $\rho_i(x)\propto \exp[-\beta {\cal H}(x,i\Delta\lambda)]$ is
the canonical distribution corresponding to the Hamiltonian ${\cal
H}(x,i\Delta\lambda)$ at step $i$, and $p_i(x \rightarrow x')$ is the
noise-averaged probability for a phase space point $x$ at the
beginning of a constant-$\lambda$ interval to evolve into phase space
point $x'$ at the end of the interval.  The caret in
Equ.~(\ref{equ:balance}) indicates {\em time reversal}, so that
$\hat{p}({x}' \rightarrow {x})$ is the probability for $x'$ to evolve
into $x$ under dynamics running backward in time.  For Langevin
dynamics, time reversal can be achieved simply by inverting the signs
of all momenta contained in phase space points $x$ and $x'$.  For a
given noise history, such transition probabilities are specified by
the deterministic map $\phi[x;\eta(t)]$.  We can thus rewrite
Equ.~(\ref{equ:balance}) as
\begin{eqnarray}
& &
\hspace{-0.5cm}
e^{-\beta {\cal H}(x,i\Delta\lambda)}
\left\langle
\delta(x'-\phi[x;\eta(t)]) 
\right\rangle_\eta
=
\nonumber\\
& &
\hspace{-0.4cm}
e^{-\beta {\cal H}(x',i\Delta\lambda)}
\left\langle
\delta(x'-\phi[x;\eta(t)])
\left|{\partial \phi[x;\eta(t)] \over \partial x}\right|
\right\rangle_\eta ,
\label{equ:balance2}
\end{eqnarray}
where angled brackets with a subscript $\eta$ denote an average 
over realizations of the noise history.  

From the condition of detailed balance, we now obtain a general
identity for averages involving the Jacobian determinant.  We begin by
multiplying Equ.~(\ref{equ:balance2}) by an arbitrary function
$f(x,x')$ of two phase space points, and then integrate over $x'$,
yielding
\begin{eqnarray}
&&
\left\langle
f(x,\phi[x;\eta(t)]) 
\right\rangle_\eta =
\left\langle
f(x,\phi[x;\eta(t)])
\right.
\nonumber\\
&&
\left.
\quad
\times
e^{-\beta [{\cal H}(\phi[x;\eta(t)],i\Delta\lambda) -
{\cal H}(x,i\Delta\lambda)]}
\left|{\partial \phi[x;\eta(t)] \over \partial x}\right|
\right\rangle_\eta.
\label{equ:balance3}
\end{eqnarray}
This result holds for any period during which the control parameter is
held constant, i.e., while no physical work is done on a system.
Since we have assumed here that equations of motion are integrated
with an infinitesimally small time step, the ``error'' work vanishes
as well.  The energy change ${\cal H}(\phi[x;\eta(t)],i\Delta\lambda)
- {\cal H}(x,i\Delta\lambda)]$ in Equ.~(\ref{equ:balance3}) is thus
identically the heat ${\cal Q}_i$ absorbed from the bath during the
$i$th time interval.  Equation~(\ref{equ:balance3}) states that
the Jacobian determinant and $e^{-\beta {\cal Q}}$ negate one
another in averages over noise history.

This proof is completed by decomposing the total change in energy
along a trajectory into contributions from physical work $W_\lambda$
(accumulated over many discrete steps in control parameter at fixed
phase space point) and heat (accumulated over many intervals during
which the phase space point evolves at fixed $\lambda$):
\begin{eqnarray}
{\cal H}(\phi[x; \eta(t)], \lambda_B)-{\cal
H}(x, \lambda_A) = W_\lambda + {\cal Q}
\\
{\cal Q} = \sum_{i=0}^{n-1}{\cal H}[x_{i+1}, i \Delta
\lambda]-{\cal H}[x_i, i \Delta \lambda]
\end{eqnarray}
We have retained the definition of physical work from
Equ.~(\ref{equ:Weps}).  Note that heat in the case of stochastic
dynamics is defined in precisely the same way as ``error'' work was
defined for deterministic dynamics propagated in an approximate way.
Recalling the generalized definition of work $W_\phi$ and applying the
identity~(\ref{equ:balance3}) with $f(x,x')=\exp\{-\beta ({\cal
H}[x', i \Delta \lambda]-{\cal H}[x', (i-1) \Delta \lambda] )\}$ for
each constant-$\lambda$ interval, we finally have
\begin{eqnarray}
\left\langle
e^{-\beta W_\phi}
\right\rangle_\eta &=& 
\left\langle
e^{-\beta W_\lambda}e^{-\beta {\cal Q}} 
\left|{\partial \phi[x;\eta(t)] \over \partial x}\right|
\right\rangle_\eta
\nonumber\\
&=& 
\left\langle
e^{-\beta W_\lambda}
\right\rangle_\eta.
\end{eqnarray}
This equivalence, together with Equ.~(\ref{equ:stochastic}), completes
an alternative route to Jarzynski's identity for systems evolving
stochastically under the constraint of detailed balance.

The consequence of detailed balance expressed in this way has
interesting implications for energy fluctuations of stochastic systems
at equilibrium (i.e., with fixed control parameter).  With the choice
$f(x,x')=1$, we have for the specific case of Langevin dynamics (see
Equ.~(\ref{equ:lang_jac}))
\begin{equation}
\left\langle
e^{-\beta {\cal Q}}
\right\rangle =
e^{n_{\rm f}\gamma\tau}
\label{equ:expQ}
\end{equation}
along trajectories of length $\tau$.  This identity stands in stark
constrast to the corresponding exponential average of energy
fluctuations under norm-conserving deterministic mappings,
Equ.~(\ref{equ:Wepsilon}).  The long-time divergence in
Equ.~(\ref{equ:expQ}) would be expected for a system which
asymptotically loses all memory of its initial conditions.  In that
case, the exponential average factorizes in the long-time limit:
\begin{equation}
\left\langle
e^{-\beta {\cal Q}}
\right\rangle \approx 
\left\langle
e^{\beta {\cal H}(x_0)}
\right\rangle
\left\langle
e^{-\beta {\cal H}(x_\tau)}
\right\rangle.
\label{equ:Qfactor}
\end{equation}
The first factor on the right hand side of Equ.~(\ref{equ:Qfactor})
averages a quantity that negates the effect of Boltzmann weighting,
and is proportional to the entire phase space volume.  Since the range
of possible momenta is unbound even in a system with finite volume, a
divergent result is inevitable once correlations have decayed
completely.  We anticipate similarly unbounded growth of exponentially
averaged energy fluctuations for many classes of stochastic dynamics,
such as Monte Carlo sampling.  That the analogous average is fixed at
unity for deterministic propagation rules with unit Jacobian such as the Verlet algorithm
indicates that errors arising from finite time step size do not
disrupt substantial correlations with initial conditions. 

\section{Conclusion}
\label{sec:conclusion}

By considering general invertible phase space mappings we have
demonstrated that the Jarzynski relation remains exactly valid
for non-equilibirum trajectories generated with large time steps,
provided the work performed on the system is defined
appropriately. For integration algorithms that conserve phase space
volume, such as the Verlet algorithm, this definition is particularly
simple.  Here, the work just equals the energy difference between the
final and the initial state of the trajectory. Simulating dynamics
with a larger time step requires fewer integration steps to
generate a trajectory of given length, and therefore lower
computational cost. Numerical simulations indicate that
optimum efficiency is achieved for time steps just short of the
stability limit. Compared to simulations with time steps of
conventional size, the long time step approach can yield improvements
in efficiency of one or more orders of magnitude.

Recently, Sun has shown how work-biased path sampling can be used to
improve the efficiency of fast switching simulations
\cite{SUN}. However, it seems that this path sampling approach does
not outperform conventional methods for calculating free energy
differences.  It will be interesting to see if the fast switching
approach can be improved, by combining the long time step
approach of this paper with biased path samping methodologies
\cite{SUN,YTREBERG,jarTPS}, to the point that it is computationally
competitive with other free energy calculation techniques, such as
umbrella sampling, thermodynamic integration, or flat histogram
sampling.

\begin{acknowledgments}
This work was supported by the Austrian Science Fund (FWF) under Grant No. P17178-N02. 
\end{acknowledgments}

\bibliographystyle{prsty}

\newpage

\end{document}